\documentclass[final,3p,times,11pt,authoryear]{elsarticle}

\usepackage{graphicx}
\usepackage{float} 
\usepackage{amssymb}
\usepackage{amsmath}
\usepackage{bbm} 
\usepackage{lineno}

\modulolinenumbers[5]
\usepackage{hyperref}
\hypersetup{colorlinks=true,citecolor=blue}
\usepackage{color}
\usepackage{multirow}
\usepackage{booktabs}
\usepackage{bm}
\usepackage{epstopdf}
\usepackage{upgreek}
\usepackage{subfigure}
\usepackage{appendix}
\usepackage{setspace}
\usepackage{amsmath}

\usepackage{framed} 
\usepackage{multicol} 
\usepackage{nomencl} 
\renewcommand*{\nompreamble}{8pt}
\makenomenclature
\renewcommand*\nompreamble{\begin{multicols}{2}}
\renewcommand*\nompostamble{\end{multicols}}
\usepackage[font=small,labelfont=bf,labelsep=space]{caption}
\usepackage{soul}

\begin{document}

\begin{frontmatter}


\title{The influence of hydrogen core force shielding on dislocation junctions in iron}




\author[materials]{Haiyang Yu}
\ead{haiyang.yu@materials.ox.ac.uk}
\author[kings,ims]{Ivaylo H. Katzarov}
\ead{ivaylo.katsarov@kcl.ac.uk}
\author[kings]{Anthony T. Paxton}
\ead{tony.paxton@kcl.ac.uk}
\author[engineering]{Alan C.F. Cocks}
\ead{alan.cocks@eng.ox.ac.uk}
\author[engineering,materials]{Edmund Tarleton \corref{cor}}
\ead{edmund.tarleton@eng.ox.ac.uk}

\cortext[cor]{Corresponding author}
\address[materials]{Department of Materials, University of Oxford, Parks Road, OX1 3PH, United Kingdom}
\address[engineering]{Department of Engineering Science, University of Oxford, Parks Road, OX1 3PJ, United Kingdom}
\address[kings]{Department of Physics, King's College London, Strand, London WC2R 2LS, United Kingdom}
\address[ims]{Bulgarian Academy of Sciences, Institute of Metal Science, 67 Shipchenski prohod, 1574 Sofia, Bulgaria}

\begin{abstract}
\textcolor{black}{The influence of hydrogen on dislocation junctions was analysed by incorporating a hydrogen dependent core force into nodal based discrete dislocation dynamics.} Hydrogen reduces the core energy of dislocations, which reduces the magnitude of the dislocation core force. We refer to this as \emph{hydrogen core force shielding}, as it is analogous to hydrogen elastic shielding but occurs at much lower hydrogen concentrations. The dislocation core energy change due to hydrogen was calibrated at the atomic scale accounting for the nonlinear inter-atomic interactions at the dislocation core, giving the model a sound physical basis. Hydrogen was found to strengthen binary junctions and promote the nucleation of dislocations from triple junctions. Simulations of microcantilever bend tests with hydrogen core force shielding showed an increase in the junction density and subsequent hardening. These simulations were performed at a small hydrogen concentration realistic for bcc iron.
\end{abstract}


\end{frontmatter}

\section{Introduction}
\label{intro}
Understanding the effect of hydrogen on dislocations \citep{Robertson2001} is essential to understanding how hydrogen reduces ductility in metals. Using in situ TEM testing, hydrogen has been observed to enhance dislocation mobility in a variety of materials \citep{Tabata1983,Robertson1986,Shin1988}. \cite{Ferreira1998} found that hydrogen decreases the equilibrium dislocation spacing in a pileup in 310 stainless steel and Al. The same authors then observed hydrogen suppresses dislocation cross slip thereby increasing slip planarity \citep{Ferreira1999}. Motivated by these experimental observations, \cite{Birnbaum1994} used analytic solutions to calculate the elastic stress field generated by a hydrogen concentration field that is in equilibrium with an edge dislocation (assuming plane strain). The elastic interaction between two dislocations or between a dislocation and a centre of dilation (defect) is reduced in the presence of hydrogen, leading to reduced spacing and enhanced mobility. This mechanism is able to partially account for the experimental observation of hydrogen enhanced plasticity, and is referred to as the hydrogen elastic shielding mechanism. Based on this mechanism, the effects of hydrogen on dislocation nucleation (Frank-Read source operation) and expansion and cross slip have been discussed \citep{Delafosse2008}. All of these studies focused  on dislocation behaviour limited to a single slip plane, and considered only the long range elastic stress, which is important for dislocation glide, but neglected the near-core contribution of hydrogen. Recently, it was revealed that hydrogen weakens dislocation junctions in fcc materials which will likely play an important role in work hardening \citep{Yu2019b}.

A sessile dislocation junction forms when a glissile dislocation cuts through a forest dislocation that does not lie on the same slip plane. This mechanism contributes to the formation of complex, tangled dislocation arrays during crystal deformation \citep{Grilli2018}, and is a key source of strain hardening. Given their importance, dislocation junctions have been extensively studied \citep{Bulatov2006}. TEM observations of dislocation structures of hydrogen-charged specimens in bcc Fe show that a homogeneous dislocation forest in a hydrogen-free sample transforms into cell walls that partition dislocation-free regions when it is charged. The cell walls can be regarded as dense dislocation tangles. TEM images reveal that the area of the dislocation free zones and the density in the dislocation tangles increase with hydrogen concentration in the interval 0-25 appm \citep{Wang2013}. The physics underlying dislocation reorganization due to hydrogen is not yet well established, but investigations into the effect of hydrogen on dislocation junctions provide a useful starting point for understanding this phenomenon.

It is instructive to observe a single dislocation junction and study its formation and destruction. This will help predict and understand the mergent properties of dislocation tangles. Due to the difficulty in pinpointing and following the development of a single junction, only a handful of experimental observations have been reported in the literature \citep{Wu2009,Ma2010,Lee2013,Pardoen2015}. This research investigated the formation and destruction of Lomer junctions using TEM, in the absence of hydrogen. Experimental investigation becomes even more challenging when hydrogen is present. Conditions apart from the hydrogen concentration ought to be equivalent between the charged and uncharged experiments, including the initial configuration of the junction, (strictly speaking) its surrounding dislocation structure and the loading history. Due to the statistical nature of dislocation activity, exact comparison between hydrogen charged and unchanged specimens is out of the question. Even for in situ tests, it is extremely difficult, if not impossible, to keep all the control variables unchanged. For instance, hydrogen charging will inevitably alter the dislocation structure, and the loading history will be distinct for different tests. No experimental observations on the effect of hydrogen on junctions have been reported for individual dislocation junctions, and it is questionable whether this will be achieved in the near future.

In the absence of suitable experiments, modelling can provide detailed and important information about the development of junctions and how they influence cell formation. Molecular dynamics (MD) \citep{Bulatov1998} and 3D discrete dislocation dynamics (DDD) \citep{Cai2006book} are powerful tools for simulating dislocation junctions. \cite{Rodney1999} employed the quasicontinuum (QC) method, which combines MD and the finite element method (FEM), to study the formation and destruction of Lomer-Cottrell junctions. \cite{Bulatov1998,Yamakov2003} simulated Lomer-Cottrell junctions using molecular dynamics simulation. Molecular statics simulation can also be used to simulate dislocation locks \citep{Parthasarathy1996,Madec2003}. Atomistic methods are limited to small temporal and spatial scales, compromising their ability to capture the long-range character of dislocation stress fields \citep{Martinez2008}; in general, the results obtained via atomistic methods cannot be directly implemented in large scale (e.g. crystal plasticity) models. DDD is an ideal tool for bridging the gap in space and time scales between atomistic and continuum models. DDD has a dislocation line segment as its basic element and uses analytic solutions for the elastic fields so is able to simulate the collective behaviour of a large array of dislocations. To date, DDD has been used to study problems ranging from individual dislocation junctions \citep{Madec2002}, to large scale plasticity arising from the motion of a large number of dislocations \citep{Zbib1998}. \cite{Shenoy2000} performed DDD simulations of a Lomer-Cottrell junction, reproducing the atomistic results obtained by \cite{Rodney1999}. \cite{Madec2002} performed systematic DDD simulations of Lomer-Cottrell junctions covering all possible initial configurations, and further evaluated the effect of forest hardening on plasticity using large scale DDD simulations. \cite{Madec2003} utilised the DDD approach in combination with molecular statics simulations to study the collinear interaction of dislocations. \cite{Bulatov2006} probed the properties of multi-junctions using DDD and \cite{Lee2011} performed DDD simulations of Lomer junctions in a free-standing thin film. \cite{Wu2016} studied binary junctions in hexagonal close-packed crystals, combining the line tension model and DD simulations.

Despite the large literature on modelling the effects of hydrogen on mechanical properties, across a wide range of scales, very few have investigated the effect of hydrogen on dislocation junctions. \cite{Hoagland1998} studied the effects of hydrogen on a Lomer-Cottrell lock using Monte Carlo (MC) calculations. \cite{Chen2008} investigated the effects of a general solute on a Lomer junction, considering different solute properties, using a kinetic Monte Carlo (kMC) coupled DDD approach. Hydrogen redistribution during the destruction of the dislocation junction was not considered in these works. Recently, \cite{Yu2019b} utilised a hydrogen informed discrete dislocation dynamics (DDD) approach to investigate the influence of hydrogen on Lomer junctions; hydrogen redistribution was taken into account and various initial configurations \citep{Madec2002} were examined. This could provide reference for modelling hydrogen with crystal plasticity approach considering dislocation interactions \citep{Gustavo2018}. \cite{Zhao2018} performed MD simulations of a hydrogen charged nanoindentation test and discussed the effect of hydrogen on dislocation entanglement.

Recently, \cite{Gu2018} proposed a framework for incorporating hydrogen into the non-singular DDD formulation \citep{Cai2006}. Hydrogen is treated as an isotropic point dilatation in an infinite elastic medium, based on which the hydrogen elastic stress field and its influence on dislocation motion are evaluated. This formulation accounts only for the linear elastic contribution of hydrogen residing outside the dislocation core region, which is consistent with the hydrogen elastic shielding mechanism. Similar formulations were also proposed in 2D by \cite{Cai2014,Song2018}. However, this is only a partial contribution of hydrogen. Neglecting the core contribution means an unrealistically high bulk concentration must be assigned in order to produce a noticeable hydrogen influence.

As described by the hydrogen elastic shielding theory, hydrogen forms atmospheres around dislocations, reducing the long-range elastic dislocation-dislocation interactions \citep{Birnbaum1994,Gu2018}. At high hydrogen concentrations, the hydrogen shielding effect increases, which could affect the formation of dislocation structures. Although the shielding effect remains commonly quoted in the current literature as a viable mechanisms for the reduction of elastic interactions between dislocations, it is not adequate to explain the experimental observations in Fe. Theoretical studies using continuum models show that the hydrogen shielding effect is relatively short range and unlikely unless hydrogen concentrations are extremely high ($> 10^{4}$ appm) \citep{Birnbaum1994}. Experiments using in-situ TEM on samples deformed in tension \citep{Ferreira1998} also show that hydrogen has a significant effect on dislocation spacing at distances greater than 20b, indicating that hydrogen shielding of the dislocation-dislocation interactions cannot, by itself, account for the observations. Atomistic simulations of the effects of hydrogen on $1/2[111]$ edge dislocation pile-ups in bcc-Fe also indicate that the shielding mechanism is not operative at hydrogen concentrations lower than $10^{5}$ appm \citep{Song2014}. Thus, other mechanisms, such as the effect of hydrogen on the dislocation core must be examined in more detail.

To account for the hydrogen core contribution, \cite{Yu2019a} implemented a hydrogen dependent dislocation mobility law where the velocity of a dislocation is increased in the presence of hydrogen. The increased mobility is attributed to hydrogen promoted kink-pair nucleation in the near core region, which was revealed using kMC simulations \citep{Katzarov2017}. Therefore, a hydrogen dependent dislocation mobility law, to some extent, accounted for the effect of hydrogen residing near the dislocation core, and allowed the effect of hydrogen on a microcantilever to be simulated at a realistic concentration. Hydrogen residing in the core  will not only affect kink-pair formation and migration but also change the dislocation core energy profile which is critical for dislocation line tension and near-field interactions. 
\cite{Cai2018} suggested introducing a solute-solute interaction energy term to the hydrogen stress formulation. In this way, an additional free energy contribution from the near core regime is incorporated, and this can be approximated as a change in the core energy due to hydrogen. With this method, \cite{Cai2018} were able to observe the effects of hydrogen on a dislocation loop and a Frank-Read (F-R) source at realistic bulk concentrations. In particular, a large decrease in the activation stress of the F-R source was observed, indicating the importance of the hydrogen core contribution in the activation of dislocation sources.

The force acting on a dislocation can be partitioned into two parts \citep{Arsenlis2007}: the elastic force and the core force, due to the dislocation core energy $E^{c}$. Hydrogen modifies this core energy and so contributes an additional hydrogen core force. The hydrogen core contribution is key in hydrogen-junction interactions, where the dislocation self force and short-range interactions are crucial. Hydrogen elastic shielding and hydrogen dependent dislocation mobility contribute marginally to dislocation junction formation which is regarded as a quasi-static process. As a demonstration, we performed DDD simulations of junction formation at low concentrations ($0-60$appm) considering these contributions and observed practically no change in junction length in the presence of hydrogen; see \autoref{subsec:application} for further details. To capture the effect of hydrogen on dislocation junctions, the hydrogen core force needs to be incorporated. The approach proposed by \cite{Cai2018} is a viable one, but as they noted, the near-core contribution was based on the stress field derived from linear elasticity theory, and the nonlinear interatomic interactions at the dislocation core were neglected. These nonlinear effects should be incorporated, which can be achieved by calibrating the dislocation core energy in the presence of hydrogen with atomistic simulations.

In this work we assume the hydrogen core force arises as a result of hydrogen lowering the dislocation core energy, which is calibrated based on first principles (DFT) calculations for bcc iron.   \textcolor{black}{We will show that the effects of hydrogen on dislocation junctions are dominated by the hydrogen core force. This highlights the critical role of dislocation core energy in near-field dislocation interactions. This work, in a sense, depicts a more general picture of how dislocation reactions are influenced by a change in dislocation core energy, using hydrogen as a medium to trigger the change.}

This paper is organised as follows: \autoref{sec:DFT} presents the details of the calibration of the hydrogen core energy using DFT. \autoref{sec:linetension} discusses the effects of the hydrogen modified core energy on dislocation junction formation in bcc iron, using a line tension model. \autoref{sec:DDD} implements a hydrogen core force in the DDD framework and employs it to study the influence of hydrogen on dislocation junction properties. A summary is presented in \autoref{sec:sum}.

\section{\textcolor{black}{Atomic calibration}}
\label{sec:DFT}
\begin{figure}[!htb]
\centering
\includegraphics[width=0.7\linewidth]{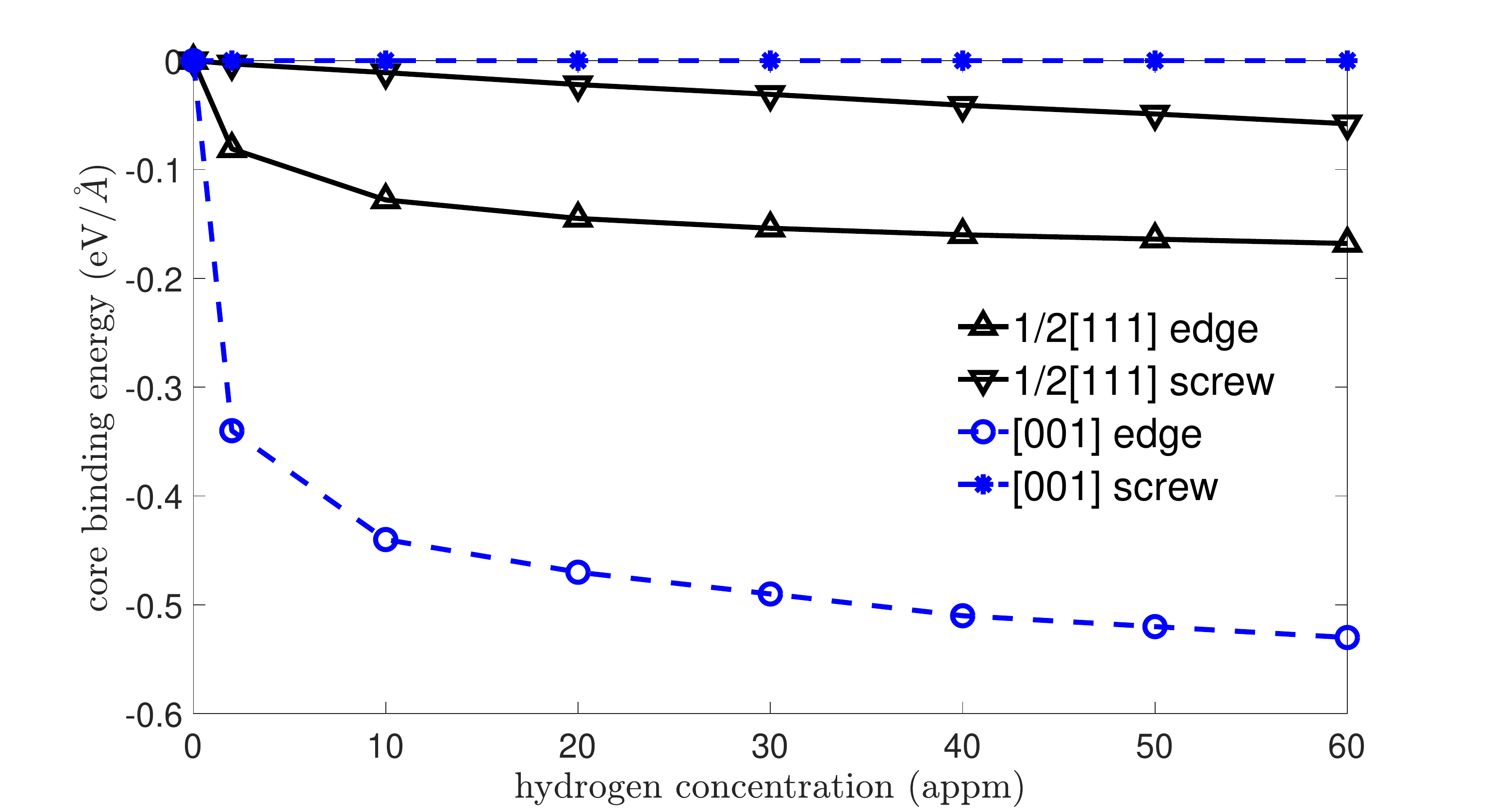}
\caption{\textcolor{black}{The calibrated hydrogen core energy for $1/2[111]$ and $[100]$ dislocations. The hydrogen core energy is negative and so will reduce the total core energy and total core force. We refer to this as \emph{hydrogen core force shielding}, as it is analogous to hydrogen elastic shielding \citep{Birnbaum1994}.} }
\label{fig:Henergy}
\end{figure}

The line energy, $E$, of a dislocation can be partitioned in two parts \citep{Arsenlis2007},
\begin{equation} \label{eq:DFT1}
E=E^{e}+E^{c},
\end{equation}
the elastic energy $E^{e}$ and the core energy $E^{c}$. $E^{e}$ can be calculated with linear elasticity theory whereas $E^{c}$ is the energy contained within the dislocation core, \textcolor{black}{typically} within $5b$ from the dislocation line. 
%
%
In an isotropic crystal the elastic energy per unit length of a straight dislocation line is
\begin{equation} \label{eq:DFT2}
E^{e}=\frac{\mu b^{2}}{4\pi(1-\nu)}\ln\left( \frac{R}{r_{c}} \right) (1-\nu \cos^{2}\theta),
\end{equation}
$\mu$ is the shear modulus, $\nu$ is Poisson's ratio,  $R$ and $r_c$ are the outer and inner cutoff radii,
and $\theta$ is the character angle between the dislocation's Burgers vector and line direction. Material parameters for iron are used in this work: $a=2.856$ \r{A} is the lattice parameter, $\mu=82$ GPa and $\nu=0.29$. Core energies in bcc Fe were determined by \cite{Clouet2009,Clouet2009b,Zhang2011} who obtained values of $E^c_{s}=0.219$ eV/\AA{} for a $1/2[111]$ screw, $E^c_{e}=0.286$ eV/\AA{} for a $1/2[111](110)$ edge, and $E^{c}_{e}=0.62$ eV/\AA{} for a $[100]$ edge dislocation, calculated with a core radius $r_{c}$, of 3 \AA{}, 2.45 \AA{} and 5.16 \AA{},
respectively.
We will notionally divide the hydrogen atoms into those which are trapped at distances greater than the core radius and usually referred to as forming an "atmosphere", and those which are occupying deep traps within the dislocation core. 
Due to the fast hydrogen diffusion compared to the low mobility of intersecting dislocations, it is reasonable to assume that
dislocations are in equilibrium with hydrogen \citep{Gu2018,Yu2019a}.
In the case of the hydrogen in the atmosphere, these are occupying sites that we will assume to be equivalent to sites in an undistorted tetrahedral interstice and since these sites are not regular tetrahedra the associated strain is not purely dilatational. 
According to calculations using the density functional theory \citep{Wang2013b}, the strain field of a single hydrogen atom in
bcc Fe is
\begin{equation}
   \bm{\varepsilon}^H = \begin{bmatrix}
      \varepsilon^H_{11} & 0 & 0 \\
      0 & \varepsilon^H_{22} & 0 \\
      0 & 0 & \varepsilon^H_{33}
    \end{bmatrix}
    = \begin{bmatrix}
      0.014 & 0 & 0 \\
      0 & 0.041 & 0 \\
      0 & 0 & 0.041
    \end{bmatrix}.
\end{equation}
The interaction energy per unit dislocation length, between a hydrogen in the atmosphere and a straight
dislocation can then be determined following the continuum approach developed by \cite{Cochardt1955}.
The interaction energy is
%
\begin{equation} \label{eq:DFT5}
E^{b}(r)=a^{3}\sum_{ij}\sigma_{ij}(r) \varepsilon^{H}_{ij}
\end{equation}
where $a$ is the lattice constant, $\sigma_{ij}$ is the stress field generated by the
straight dislocation
expressed in the same coordinate system as $\bm{\varepsilon}^{H}$.
\textcolor{black}{The total interaction energy per unit dislocation length between the hydrogen atmosphere and a straight
dislocation is
\begin{equation}\label{eq:continuum}
E^{H}(\chi_{0}) = \int_{0}^{2\pi}\int_{r_{H}}^{R} C_{max} \chi(\varphi, r) E^{b}(\varphi, r) r d\varphi dr
\end{equation}
where $r_{H}$ and $R_{H}$ are the inner and  outer cut-off radii of the hydrogen atmosphere centred at the
dislocation. $C_{max}$ is the maximum solute concentration for the solid (in solutes per unit volume), and $\chi\equiv C/C_{max}$ 
is the fraction of available lattice sites that are occupied by hydrogen \citep{Cai2018}.} 
%
Turning now to the hydrogen atoms trapped within the core,
their binding energies, $E^b$, can be calculated using atomistic simulations
\citep{Kimizuka2011}. These show that $1/2[111](110)$ edge dislocations
generate seven inequivalent hydrogen trap sites in the plane perpendicular to the dislocation line sense, with binding energy higher
than 100 meV (one with binding energy $E^{b}_{1}=0.34~\textrm{eV}$, two with binding energies $E^{b}_{2}=0.40~\textrm{eV}$,
$E^{b}_{3}=0.27~\textrm{eV}$, $E^{b}_{4}=0.12~\textrm{eV}$).
The contribution to total hydrogen--dislocation interaction energy per periodic distance along the
dislocation line from the deep near-core hydrogen binding sites is
\begin{equation} \label{eq:atomic}
G^{H}(\chi_{0}) = \sum_{i}\chi^{b}_{i}(\chi_{0}) E^{b}_{i}
\end{equation}
where the sum is over the binding sites $i$ \citep{Kimizuka2011}. 
The probability that a trap site is occupied by hydrogen is calculated from the McLean isotherm,
\begin{equation} \label{eq:DFT4}
\chi^{b}_{i} = \frac{\frac{1}{6}\chi_{0}\exp \left( E^{b}_{i}/k_{B}T \right)}{1+\frac{1}{6}\chi_{0} \exp\left( E^{b}_{i}/k_{B}T \right)}.
\end{equation}
where the factor $1/6$ accounts for there being six tetrahedral sites per iron atom in good crystal. $\chi_{0}$ is the nominal number of hydrogen atoms per iron atom: the concentration of hydrogen in atomic parts per million (appm) is thereby $c_0=10^{6}\chi_0$.
The change in energy per unit length, $E^{H}(\chi_{0})$, is then
\begin{equation}
    E^H = \frac{G^H(\chi_{0}) }{L}, \quad
    L=\left\{\begin{array}{cc}
    \sqrt{6}a & \textrm{edge}\\
    \sqrt{3}a/2 & \textrm{screw}
    \end{array}\right.
\end{equation}
where L is the periodic spacing along the dislocation line
The hydrogen concentration that we use in this work will range between zero and 60~appm; typical amounts found in industrial contexts are 1--10~appm, whereas specimens charged in the laboratory may reach 100~appm. 
In this range, the contribution to the total hydrogen--dislocation interaction energy from the near-core deep binding sites is dominant. For instance, for a $1/2[111](110)$ edge dislocation with $c_{0}=30$~appm the contribution to the total binding energy per unit dislocation length due to "core" hydrogen calculated from \autoref{eq:atomic} is 0.153~eV/\AA\,
The contributions from the hydrogen atmosphere 
calculated from \autoref{eq:continuum} with inner cut-off radius $r_{H}=a$ and outer cut-off radii $R_{H}=100a$ and $R_{H}=1000a$ are correspondingly 0.0018 eV/\AA\,and 0.0022 eV/\AA. 
In the case of the $1/2[111]$ screw dislocation, we use the deep near-core hydrogen binding sites calculated by \cite{Itakura2013}. Because the strain field is purely deviatoric in linear elasticity the change in energy due to the hydrogen atmosphere is yet smaller than for the edge dislocation. When it comes to the hydrogen binding to $[100]$ dislocations we consider only the atmospheric hydrogen and neglect core binding.
The interaction energies, $E^{H}$, as functions of nominal hydrogen concentration, $c_0$ are shown in \autoref{fig:Henergy}.
The hydrogen core energy is negative and so will reduce the total core energy and total core force. This trend is consistent with the defactant theory proposed by \cite{Kirchheim2007,Kirchheim2010}. It is noted that the hydrogen binding energy for a $1/2[111]$ screw dislocation is non-zero. This is because hydrogen has a tetragonal misfit strain which couples with the shear field of the 1/2[111] screw dislocation to produce non-zero interaction energy. 
In contrast, the hydrogen binding energy for a $[100]$ screw dislocation is zero, because the misfit strain of hydrogen does not couple with the elastic field in this case. When two dislocation segments attract and intersect to form a junction, the  dislocation lines rotate, resulting in a change in the angles between the Burgers vectors and line directions (the character angle, $\theta$). The change in the character angle leads to rearrangement of hydrogen atoms around the dislocation which changes the interaction energy. The continuum method (\autoref{eq:continuum}) can be used to calculate the interaction energy between the hydrogen atmosphere and a mixed straight dislocation. However, the comparison of $E^{H}$ calculated by the atomistic (\autoref{eq:atomic}) and continuum methods shows that $r_{H}$ for $1/2[111]$ screw and edge dislocations are different, which indicates that the inner cut-off radius depends on the character angle. Therefore, accurate determination of the interaction energy between hydrogen and a mixed dislocation is not feasible with the continuum approach without using reliable data for the inner cut-off radius, $r_{H}$, as a function of the character angle. In this study, we approximate the interaction energy $E(\theta)$ between a hydrogen atmosphere and a mixed $1/2[111]$ dislocation segment with a character angle $\theta$ as an interpolation between the interaction energies $E_H^{s}$ and $E_H^{e}$ for the screw and edge segments calculated atomistically. Inspired by the observation in \cite{Cai2018} that $E_H$ seems to have a roughly sinusoidal dependence on $\theta$, the following interpolation function is used
\begin{equation}
\label{eq:HcoreEinter0}
E^{H} = E^{H}_e\sin^2\theta + E^{H}_s\cos^2\theta,
\end{equation}
Note that this interpolation is not unique, any number of schemes which enable a smooth transition from $E^H(0)=E^H_s$ to $E^H(\pi/2)=E^H_e$ could be used. We tested other interpolation schemes, including a linear interpolation, in the line tension model, and found that the influence of the interpolation scheme on dislocation junction formation is limited. 

\textcolor{black}{Finally, it is acknowledged that the atomic calibration presented in this section is based on certain simplifications. For instance, the continuum approach is used for the $[100]$ case and a simple interpolation is applied to mixed dislocations. The key features of the hydrogen binding energy profile are captured: the binding energy is small but non-zero for a $1/2[111]$ screw and zero for a $[100]$ screw; the magnitude of the binding energy is higher for edge than screw dislocations. The main goal of this work is to present the first discrete dislocation dynamics framework utilizing atomistic-level hydrogen energetics for the dislocation core as input and investigate any emergent influence on plasticity. The outcomes of this study are more qualitative and conceptual than quantitative.}
\section{Line tension model}
\label{sec:linetension}
To investigate the role of hydrogen trapped in the dislocation core on the formation of dislocation junctions, we investigate the intersection between straight dislocations in Fe moving on different glide planes. Our aim here is to evaluate hydrogen core effects on dislocation reactions, and ignore the external stress and elastic interactions between dislocations. In bcc materials, both binary and triple junctions are important in plasticity \citep{Bulatov2006}. Binary junctions result from the interaction of two dislocations on different slip systems, and can influence the yield stress and early stages of work hardening, whereas triple junctions result from the interaction of three dislocations, and are important during the late stages of work hardening. Triple junctions are strong anchors for dislocation entanglements and are potentially important in the formation of dislocation cell structures \citep{Bulatov2006}.

\begin{figure}[ht]
\centering\includegraphics[width=0.8\linewidth]{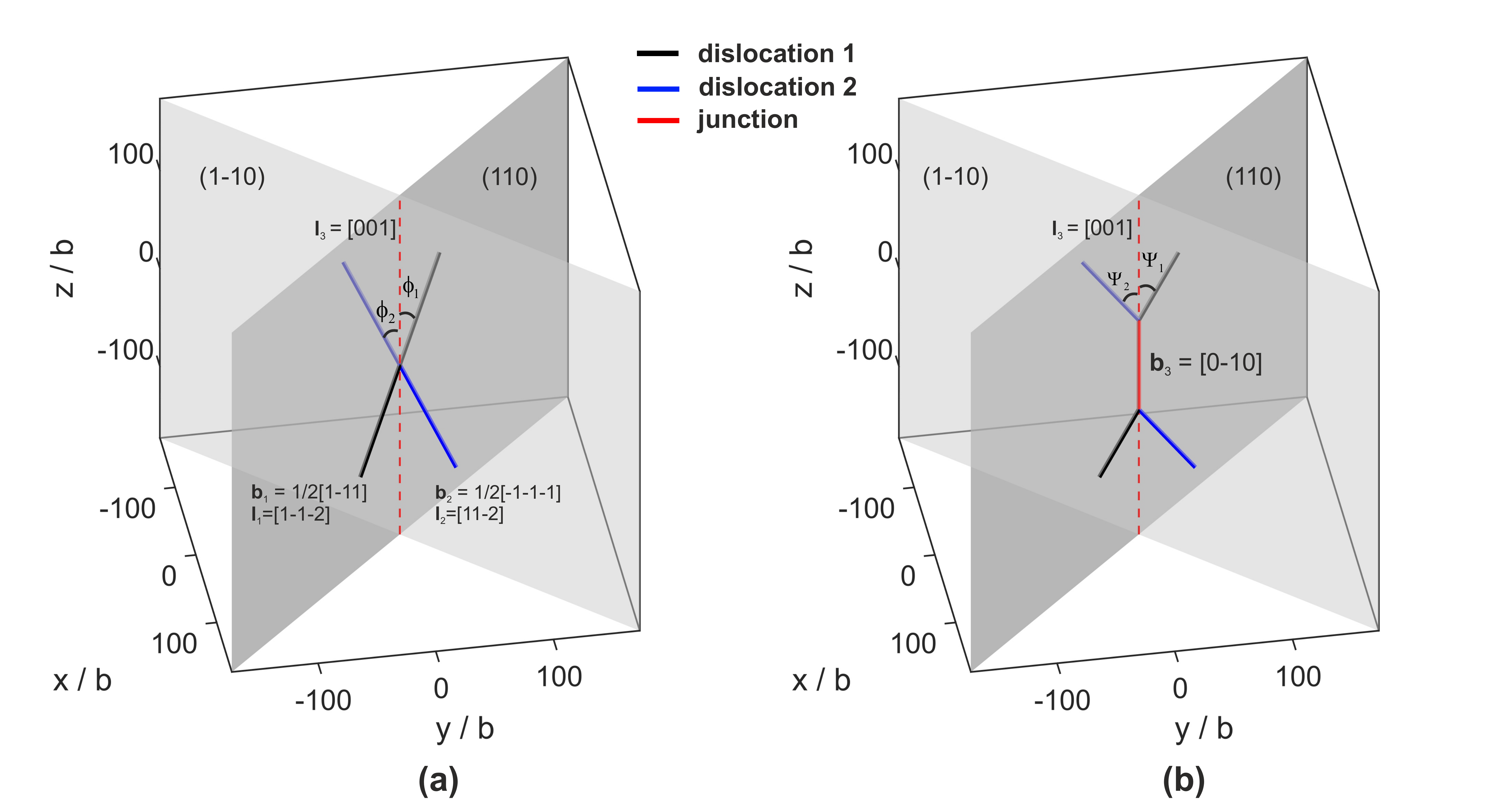}
\caption{Illustration of the binary junction geometry: (a) the initial configuration and (b) the junction formed after relaxation. The initial dislocations and the junction are of pure edge type.}
\label{fig:biillustration}
\end{figure}

The binary junction geometry is illustrated in \autoref{fig:biillustration}. Consider two straight dislocation line segments, with Burgers vectors $b_{1}$ and $b_{2}$. Both segments are bound by a pair of pinning points.
The dislocation segments are in the slip planes with normals $\bm{n}_{1}$ and $\bm{n}_{2}$ and intersect at their mid-point; which is on the line of intersection of the two slip planes, $\bm{L}=\bm{n}_2\times\bm{n}_1=[001]$. The dislocations are at angles $\phi_1 $ and $\phi_2 $ to $\bm{L}$. A junction (a straight dislocation segment along $\bm{L}$) with a Burgers vector $\bm{b}_{3}=\bm{b}_{1}+\bm{b}_{2}$ will form if it reduces the energy of the system. If $E_{1}$, $E_{2}$ and $E_{3}$ are respectively the line energies of the reacting dislocations and junction, a junction will form if the following criterion is fulfilled \citep{Baird1965}
\begin{equation} \label{eq:linetension1}
E_1\cos\phi_1+E_2\cos\phi_2 > E_{3}
\end{equation}
The energy of the configuration is a minimum when the condition
\begin{equation} \label{eq:linetension2}
E_1\cos\psi_1+E_2\cos\psi_2 - E_{3} = 0
\end{equation}
is satisfied with equilibrium angles $\psi_1 $ and $\psi_2 $, as illustrated in \autoref{fig:biillustration}(b).

\begin{figure}[!ht]
\centering\includegraphics[width=0.7\linewidth]{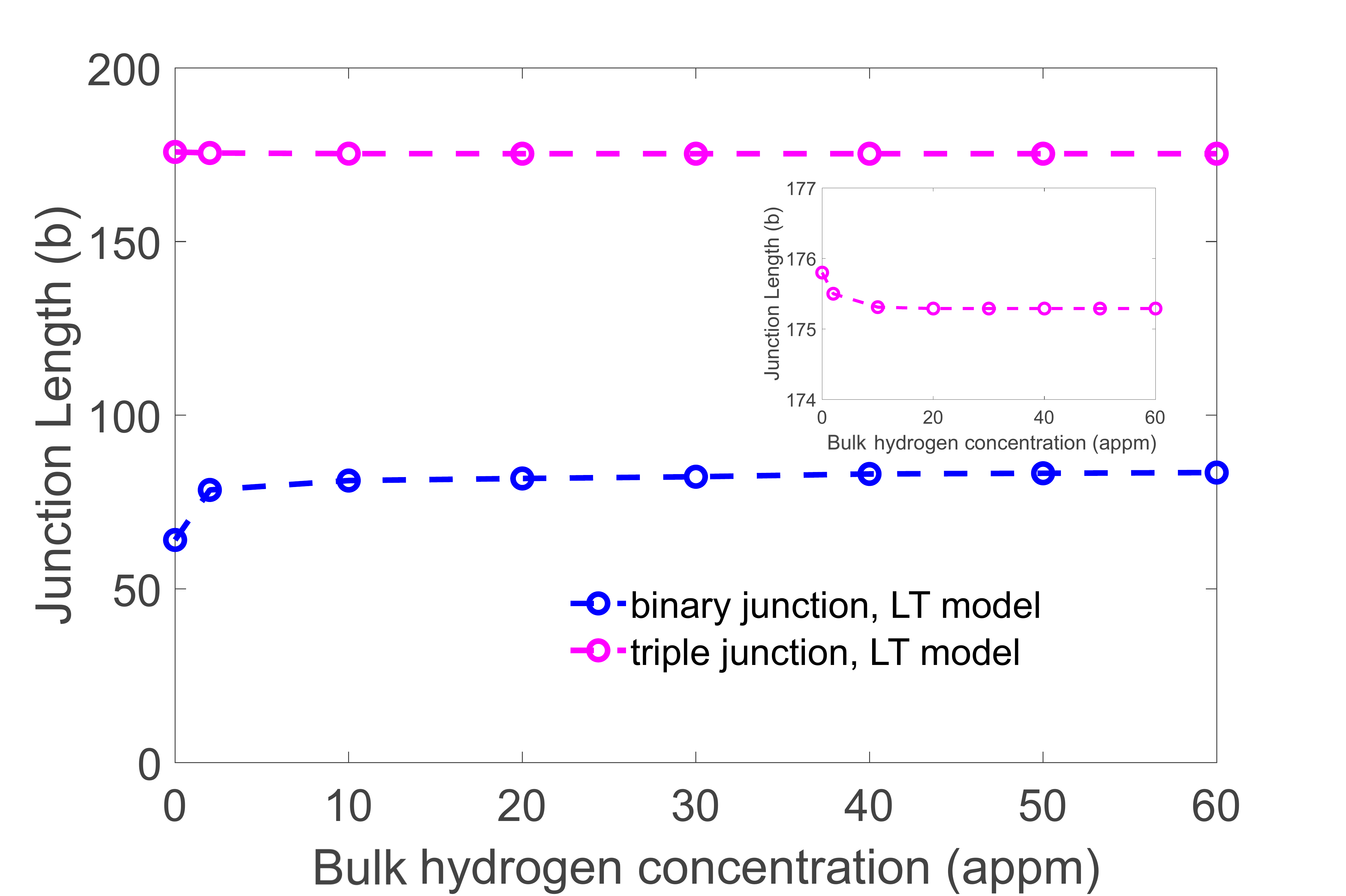}
\caption{Variation of junction length with hydrogen concentration calculated via the LT model.}
\label{fig:LTresults}
\end{figure}

In order to evaluate the effect of hydrogen on junction formation we consider the interaction between two $1/2\langle111\rangle$ edge dislocations, namely,  $1/2[1\bar{1}1](110)$ and $1/2[\bar{1}\bar{1}\bar{1}](1\bar{1}0)$, as illustrated in \autoref{fig:biillustration}.
The dislocations are of length $l_1=l_2=200b$, and initially intersect at their mid point. The dislocations attract each other and their intersection results in an $[0\bar{1}0](\bar{1}00)$ edge junction:
\begin{equation} \label{eq:linetension4}
1/2[1\bar{1}1](110)+1/2[\bar{1}\bar{1}\bar{1}](1\bar{1}0)\rightarrow [0\bar{1}0](\bar{1}00).
\end{equation}
After including the dislocation core energy change due to hydrogen $E^{H}$ (shown in \autoref{fig:Henergy}) into \autoref{eq:DFT1}, the energy per unit dislocation length in the presence of hydrogen is
\begin{equation} 
\label{eq:linetension5}
E=E^{e}+E^{c}+E^{H}.
\end{equation}
To evaluate the effect of hydrogen on the dislocation reaction, at different bulk hydrogen concentrations, we determine the junction length which minimises the energy of the configuration; by substituting the energies determined in \autoref{sec:DFT} into \autoref{eq:linetension2}. The calculations show an increase in binary junction length but almost no change in the triple junction length due to hydrogen. Since a $\langle100\rangle$ edge dislocation generates deeper binding sites (compared to a $1/2\langle111\rangle$ edge segment), the junction generated by the reaction traps more hydrogen atoms, reducing the energy of the system, and increasing the equilibrium binary junction length, as shown in \autoref{fig:LTresults}.

In the triple junction case we adopt the same configuration as in \cite{Bulatov2006}. Three $\langle111\rangle\{1\bar{1}0\}$ edge dislocations 
intersect at their midpoint and a $1/2[111]$ screw type triple junction is formed,
\begin{equation} \label{eq:triple}
1/2[\bar{1}11](01\bar{1})+1/2[1\bar{1}1](10\bar{1})+1/2[11\bar{1}](1\bar{1}0)\rightarrow 1/2[111].
\end{equation}
In this case, the screw dislocation does not generate additional binding sites and hydrogen atoms released during the reduction of the lengths of the reacting edge dislocations increases the energy of the configuration, resulting in a very slight reduction in the screw junction length with hydrogen, as shown in \autoref{fig:LTresults}. The effect of hydrogen in this case is a second order effect since the triple junction is screw type, its core energy is independent of hydrogen. It is only the edge components of the reacting dislocations (rotated $\langle111\rangle$ screw dislocations) which have a reduced core energy. This increases the equilibrium length of the reacting dislocations and therefore reduces the equilibrium length of the triple junction.

The line tension model is a static approach for evaluating hydrogen core effects on dislocation reactions. It is applicable only for straight dislocation segments, while the external stress and dislocation interactions are ignored. In reality, under an applied stress, the dislocation segments will bow out. To study the effects of hydrogen on the intersection of curved dislocations requires including both the long-range elastic interaction due to hydrogen and the short range interaction between dislocations and hydrogen at the core. Discrete dislocation dynamics (DDD) simulations on the effects of hydrogen on dislocation junctions are performed subsequently for this purpose. Furthermore, DDD can be used to determine the junction strength, which is not possible with the line tension model.
\section{DDD simulation with hydrogen core force}
\label{sec:DDD}
Hydrogen influences dislocation plasticity by exerting elastic shielding, core force shielding and modifying the dislocation mobility. Long-range hydrogen elastic shielding and a hydrogen dependent mobility law were implemented by \cite{Yu2019a}. However, the model was not able to capture the effect of hydrogen on dislocation junction formation, due to the short-range and quasi static nature of the process. To some extent, this was implied in our recent work on the influence of hydrogen on Lomer junctions \citep{Yu2019b}, where hydrogen effects were observed with an unrealistically high bulk concentration due to the omission of hydrogen core effects. DDD simulations of binary and triple junctions are performed here, considering only hydrogen elastic shielding and a hydrogen dependent mobility. The initial dislocation configurations are the same as for the line tension model in \autoref{sec:linetension}.
\begin{figure}[!ht]
\centering\includegraphics[width=0.9\linewidth]{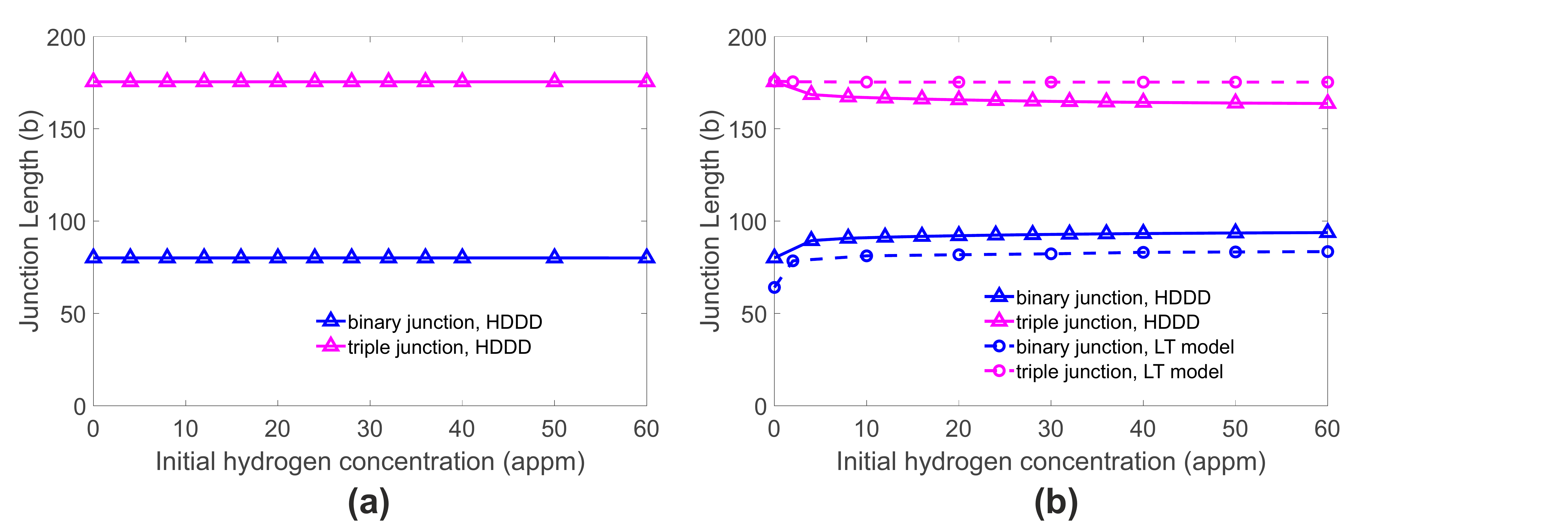}
\caption{(a) Variation of junction length with hydrogen concentration calculated via DDD without considering hydrogen core force shielding and (b) with hydrogen core force shielding.}
\label{fig:DDDlengths}
\end{figure}
As shown in \autoref{fig:DDDlengths}(a), the effect of hydrogen on junction length is not captured at these low concentrations. Comparison to the line tension model results (\autoref{fig:LTresults}) indicates that hydrogen core force shielding is dominant in the influence of hydrogen on junction formation.

\subsection{Formulation}
\label{subsec:formulation}
In the DDD simulation, dislocations are represented by discrete straight segments. The dislocation nodal velocity $\bm{V}_k$ is determined by balancing the drag force with the nodal driving force $\bm{F}_k$ on node $k$ at position $\bm{X}_k$. The nodal force is evaluated based on the non-singular continuum theory of dislocations developed by \cite{Cai2006}. In general, it consists of five parts in the presence of hydrogen
\begin{equation}
\label{eq:Fk}
\begin{aligned}
\bm{F}_{k} & = \sum_{l}\sum_{i,j}\tilde{\bm{f}}^{ij}_{kl}(\bm{X}_k) + \sum_{l}\sum_{i,j}\bar{\bm{f}}^{ij}_{kl}(\bm{X}_k) + \sum_{l}\bm{f}^c_{kl}(\bm{X}_k)+\sum_{l}\bm{f}^{c,H}_{kl}(\bm{X}_k)+\sum_{l}\hat{\bm{f}}_{kl}(\bm{X}_k)\\
& = \tilde{\bm{F}}_k+\bar{\bm{F}}_k+\bm{F}^c_k+\bm{F}^{c,H}_k+\hat{\bm{F}}_k,
\end{aligned}
\end{equation}
where $\tilde{\bm{f}}^{ij}_{kl}(\bm{X}_k)$ is the interaction force at node $k$, due to segment $i\rightarrow j$ integrated along segment $k\rightarrow l$; this is summed over all segments $i\rightarrow j$ inside the domain, including the self force due to segment $k\rightarrow l$, this is then summed over all nodes $l$ which are connected to node $k$. Similarly, $\bar{\bm{f}}^{ij}_{kl}(\bm{X}_k)$ is the hydrogen elastic shielding force evaluated at node $k$, implemented following the formulation proposed by \cite{Gu2018}. $\bm{f}^c_{kl}(\bm{X}_k)$ is the dislocation core force and $\bm{f}^{c,H}_{kl}(\bm{X}_k)$ is the hydrogen core shielding force (that is missing in previous simulations). $\hat{\bm{f}}_{kl}(\bm{X}_k)$ is the corrective elastic force for finite boundary value problem, which is evaluated with the finite element method using the superposition principle \citep{Needleman1995}.

In the presence of hydrogen, the core energy is
\begin{equation}
\label{eq:coreE}
E^{c}_\textrm{eff}(\theta) = E^{c}(\theta) + E^H(\theta)=  \sin^2(\theta) (E^{c}_e+E^{H}_e) + \cos^2(\theta) (E^{c}_s+E^{H}_s),
\end{equation}
\textcolor{black}{where $E^{c}_e$ and $E^{c}_s$ are the edge and screw core energies in the absence of hydrogen and $E^H_e$ and $E^H_s$ are the hydrogen core binding energies presented in \autoref{fig:Henergy}.} The core force \citep{Cai2018} in the presence of hydrogen is then
\begin{equation}
\label{eq:coreF}
\bm{f}^c_{kl}(\bm{X}_k)+\bm{f}^{c,H}_{kl}(\bm{X}_k) = -E^{c}_\textrm{eff}(\theta)\bm{l}_{kl} + \frac{dE^{c}_\textrm{eff}(\theta)}{d\theta}\left(\frac{\bm{b}^e_{kl}}{ \lvert\bm{b}^e_{kl}\rvert} \right),
\end{equation}
where $\theta$ is the angle between $\bm{l}_{kl}$ the line direction and Burgers vector of segment $k\rightarrow l$ with length $L_{kl}$. $\bm{b}^e_{kl}$ is the edge component of the Burgers vector. The first term on the right hand side acts along the line direction to shrink the segment, while the second term is a moment tending to rotate the segment to the orientation with lowest core energy \citep{Arsenlis2007}; in the absence of hydrogen this is the screw orientation. \textcolor{black}{As discussed in \autoref{sec:DFT}, different dislocation core radii and hydrogen free core energies were used for the $1/2[111]$ and $[100]$ dislocations in the atomistic calibration. To be consistent, we detected the dislocation type during the DDD simulation and assigned the core radii and hydrogen free core energies accordingly. For instance, for $1/2[111]$ dislocations, a core radius of $3\AA$ was used, and a core radius of $5.16\AA$ was used for $[100]$ dislocations. The hydrogen free core energies for $1/2[111]$ edge and screw dislocations are respectively $E^{c}_e=0.286$ eV/\AA{} and $E^{c}_s=0.219$ eV/\AA{} \citep{Clouet2009,Clouet2009b}; for $[100]$ edge dislocation the value is $E^{c}_e=0.62$ eV/\AA{} \citep{Zhang2011}. To the best of our knowledge, the dislocation core energy for $[100]$ screw dislocation in iron has not been reported in the literature, therefore, the core energy for this case is approximated following the relation $E^{c}_s=(1-\nu)E^{c}_e=0.44$ eV/\AA{}. It is noted that the atomistically determined core energies in the $1/2[111]$ cases also follow this relation ($0.219/0.286=0.766\approx 1-\nu$). The validity of the treatment of dislocation core radii and energies is verified in \autoref{fig:DDDlengths}(b), where good agreement is achieved in predicting the hydrogen free junction lengths using two different mthods; the line tension model and DDD.}

A linear dislocation mobility law \citep{Arsenlis2007,Yu2019a} is used. For each segment $kl$, a drag tensor $\bm{B}_{kl}$ is determined according to the segment character, the nodal velocity $\bm{V}_k$ at node $k$ is then obtained using
\begin{equation}
\label{eq:linearmob}
\left[\frac{1}{2}\sum_{l}L_{kl}\bm{B}_{kl}\right]^{-1}\bm{F}_{k} = \bm{V}_k,
\end{equation}
where the sum is over all nodes $l$ connected to node $k$, and $\bm{F}_k$ is the nodal force determined with \autoref{eq:Fk}. The mobility of a dislocation segment is anisotropic with respect to glide and climb and line directions,
\begin{equation}
\label{eq:dragedge}
\bm{B}_{kl}(\bm{l}_{kl}) = B_{g}(\bm{m}_{kl}\otimes\bm{m}_{kl})+B_{c}(\bm{n}_{kl}\otimes\bm{n}_{kl})+B_{l}(\bm{l}_{kl}\otimes\bm{l}_{kl}),\quad \mathrm{when}\ \bm{l}_{kl}\perp\bm{b}_{kl},
\end{equation}
where $B_{g}$ and $B_{c}$ are the drag coefficients for glide and climb, respectively. The unit vectors are the plane normal $\bm{n}_{kl}$ and glide direction $\bm{m}_{kl}$. 
The dislocation mobility is inversely proportional to the drag coefficient. In bcc materials, the mobility of a pure edge segment should be greater than a pure screw segment, which is accounted for by assigning $B_{eg} < B_s$. In the absence of hydrogen, the edge and screw drag coefficients are $B_{eg}=5\times10^{-4}~\textrm{Pa s}$ and $B_{s}=1\times10^{-2}~\textrm{Pa s}$ as used by \citet{Wang2011}.
The glide drag coefficient for a mixed segment is determined by interpolation between the screw and edge values using
\begin{equation}
\label{eq:Bkl}
B_g = \left[B^{-2}_{eg}\sin^2(\theta)+B^{-2}_{s}\cos^2(\theta)\right]^{-1/2}
\end{equation}
the drag coefficient for climb is set sufficiently high, $B_c=1\times10^{6}B_{eg}$ so that only glide occurs and the line drag coefficient is sufficiently low, $B_l=1\times10^{-4}B_{eg}$, to allow nodes to move freely along the dislocation line. 

Cross slip of screw segments is considered in a very simplified manner. For a screw segment, all possible cross slip planes are checked and the plane with the maximum resolved shear stress is used as the cross slip plane. The motion of the screw segment is then limited to this plane. For example, in the triple junction case, a $1/2[111]$ screw junction is formed, whose possible cross slip planes are the $(\bar{1}10)$, $(\bar{1}01)$ and $(0\bar{1}1)$ planes. Under uniaxial tension applied along $[100]$, the screw junction will cross slip on the $(\bar{1}10)$ plane.

\subsection{DDD modeling of dislocation junctions}
\label{subsec:application}
\subsubsection{Binary junction}
\label{subsubsec:binary}
The hydrogen elastic shielding and hydrogen dependent mobility law were found to be unimportant over the range of concentrations considered here. Therefore, only hydrogen core force shielding is implemented unless stated otherwise.

Using the binary junction model in \autoref{sec:linetension}, the effect of hydrogen on junction length is studied with DDD. The results are plotted against the line tension model predictions in \autoref{fig:DDDlengths}(b). In contrast to \autoref{fig:DDDlengths}(a), a significant hydrogen effect is captured, indicating the dominant role of hydrogen core force shielding in dislocation junctions. Good agreement is achieved between the DDD and line tension model. Some discrepancy is expected, as DDD accounts for the elastic interactions between segments, which is neglected in the line tension calculations.

\begin{figure}[!ht]
\centering\includegraphics[width=0.9\linewidth]{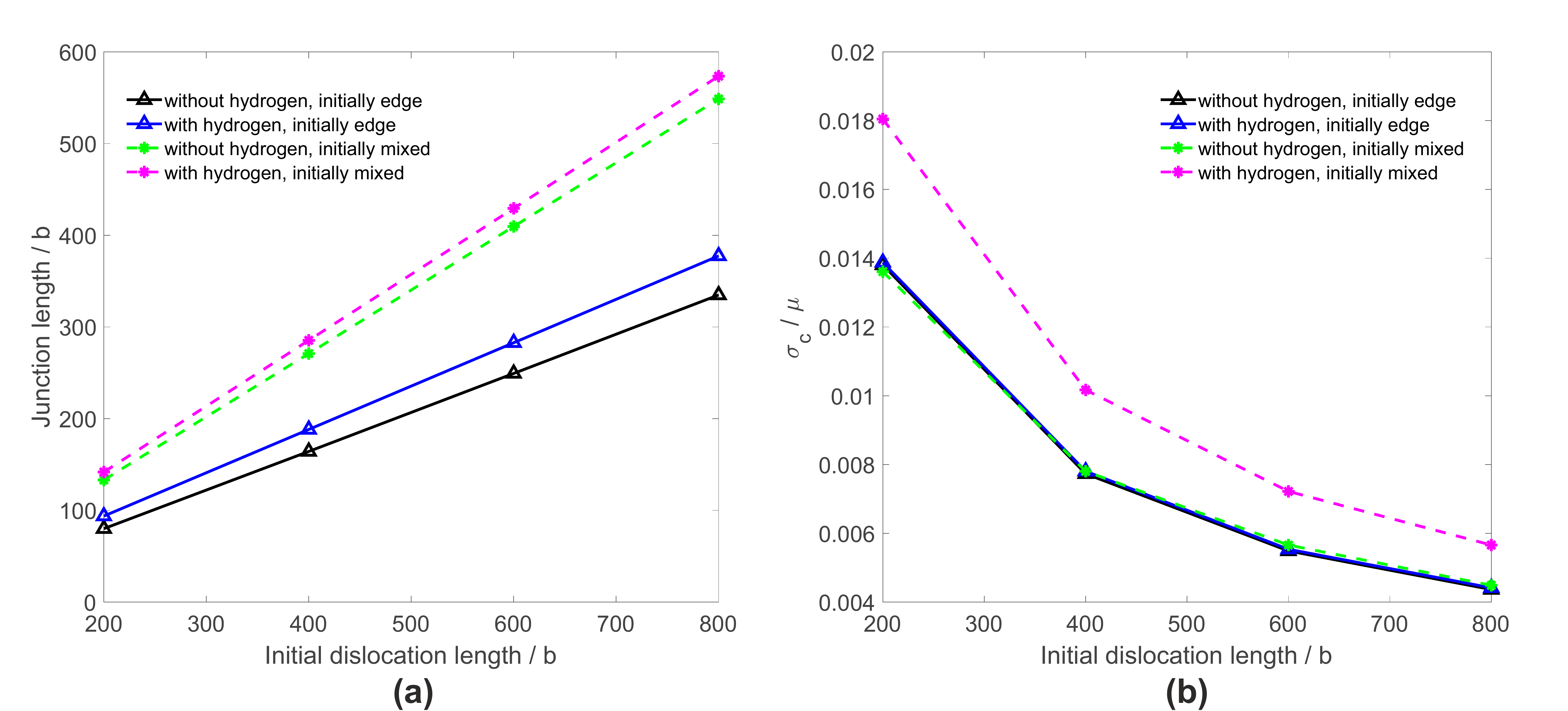}
\caption{The effect of hydrogen on binary junction (a) length and (b) strength.}
\label{fig:DDDbistrength}
\end{figure}

A binary junction can be destroyed under an applied stress. The non-junction segments can bow out, unzipping the junction, and restoring the initial configuration. 
After a sufficient relaxation time, we apply a uniaxial tensile stress along $[100]$.
%
%
As the magnitude of the stress gradually increases, the junction will be unzipped reducing the junction length. We take the stress $\sigma_c$ at which the junction length first becomes zero as the strength of the binary junction. We simulate a hydrogen concentration of $c_0=10.0$appm. The initial dislocation length is varied from $l_0=200b$ to $l_0=800b$. The junction length and strength are shown in \autoref{fig:DDDbistrength}.
Increasing the initial dislocation length, increases the junction length (to obtain the equilibrium angles $\Psi_1=\Psi_2\approx 53.5^\circ$ without hydrogen and $\Psi_1=\Psi_2\approx 56.5^\circ$ with hydrogen) and reduces the junction strength. Consistent with the observation in \autoref{fig:DDDlengths}, hydrogen increases the junction length, independently of the initial dislocation length. Hydrogen increases the strength of the junction for mixed initial segments and with initial edge segments the effect is negligible. The effect of hydrogen is dependent on the initial orientation of the intersecting dislocations, even if the line direction of the junction is fixed \citep{Yu2019b}. 
To be systematic, we rotated the initial dislocations inside their slip plane, changing their angles ($\phi_1$,$\phi_2$) with respected to the line direction of the junction ($[001]$); we tested all the cases with $\phi_1=\phi_2$ and found that hydrogen increases the strength of the $[001]$ junctions. A typical example with initially mixed segments at $\phi_1=\phi_2=20^\circ$ is shown in \autoref{fig:DDDbistrength}.

It might be assumed that hydrogen increases the junction strength by increasing the junction length, thus shortening the initial segments; making it more difficult for them to bow out and unzip the junction. However, in \autoref{fig:DDDbistrength}(a), a large difference in junction length is observed between the two initial configurations in the absence of hydrogen; in \autoref{fig:DDDbistrength}(b), however, the corresponding junction strengths are identical. This indicates the increase in junction strength due to hydrogen is not attributed to the hydrogen increased junction length but to some other mechanism.

\begin{figure}[!ht]
\centering\includegraphics[width=0.9\linewidth]{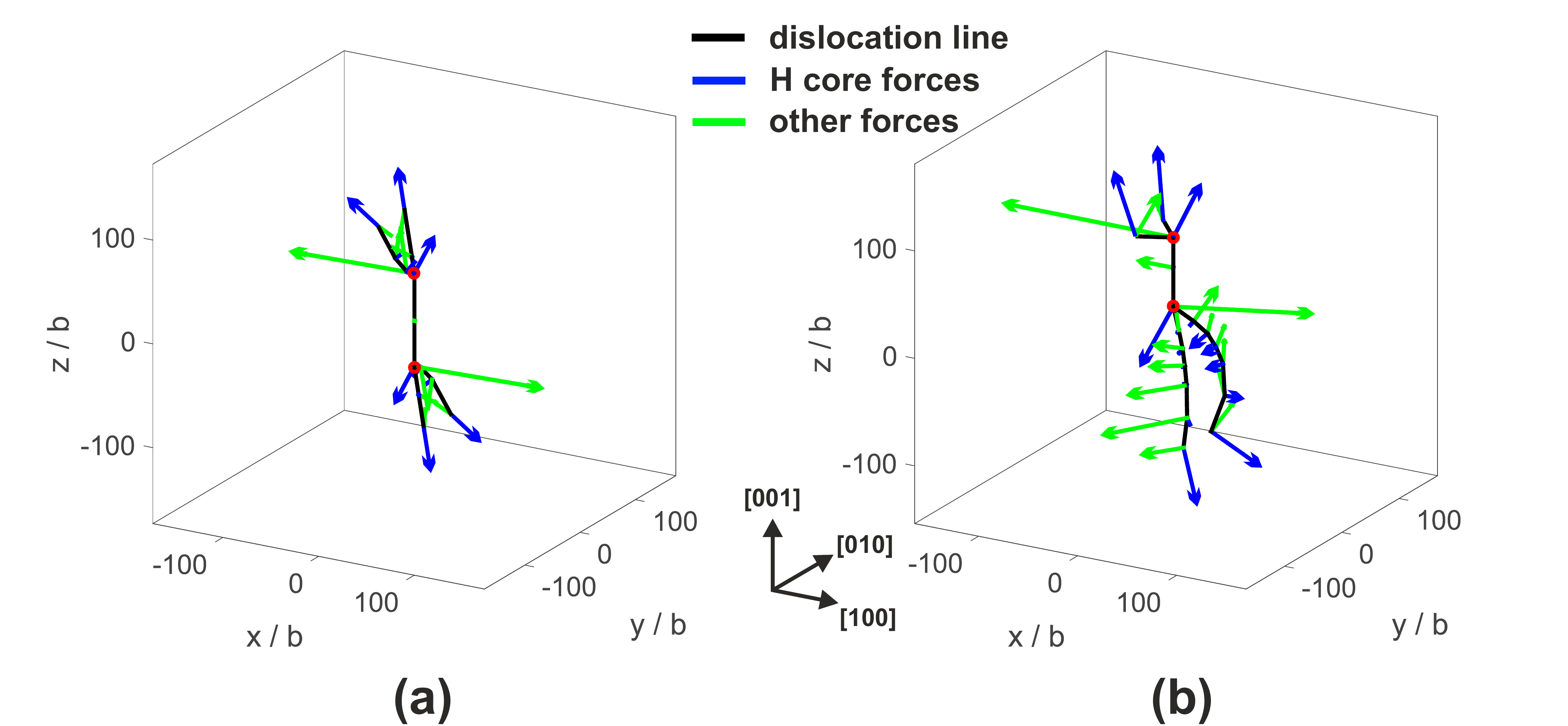}
\caption{The nodal force partitioned into hydrogen, $\bm{F}^{c,H}$ (blue arrows), and non-hydrogen parts $\tilde{\bm{F}}_k + \hat{\bm{F}}_k + \bm{F}^{c}$ (green arrows). (a) shows the formation (zipping) of a $[001]$ junction. (b) shows the destruction (unzipping) of the junction. The end points of the junction are marked with red circles;  note ${\bm{F}}^{c,H}$ has been scaled $5\times$ for clarity.}
\label{fig:biforce}
\end{figure}

The driving force for dislocation motion is partitioned into the hydrogen core force shielding term, $\bm{F}^{c,H}$, and the hydrogen free forces, $\tilde{\bm{F}}_k + \hat{\bm{F}}_k + \bm{F}^{c}$, as shown in \autoref{fig:biforce}.
During formation, hydrogen lowers the core energy of the $\langle100\rangle$ junction. The hydrogen core forces act to stretch the junction, (reducing the total core force) and increasing the junction length. The destruction of the junction is realised via a procedure similar to the operation of a F-R source. A $1/2[111]$ arm which is pinned at one end and constrained to move along the line of the junction at the other end is activated and bows out, which drives the mobile node along the line, unzipping the junction. The hydrogen core force tends to impede the expansion of this F-R source making the destruction of the junction more difficult. Finally, it should be noted that the opposite was observed in fcc materials where hydrogen elastic shielding weakens Lomer junctions \citep{Yu2019b}.

\subsubsection{Triple junction}
\label{subsubsec:triple}
Triple junctions are stronger than binary junctions. \cite{Bulatov2006} observed more pronounced strain hardening in the presence of triple junctions and attributed the phenomenon to the propensity of these junctions to form additional dislocation sources. Triple junctions occur at a late stage of loading, serving as strong anchors which form complex dislocation networks which impede dislocation motion. Unlike the $\langle001\rangle$ binary junctions which are sessile, the $1/2\langle111\rangle$ triple junctions are glissile and considerably longer than the other $1/2\langle111\rangle$ arms. As a result, the junction (rather than the non-junction segments), will bow out under an external stress, meaning rather than unzipping, the junction operates like a F-R source.

Therefore, the strength of a triple junction is defined here as the stress required to nucleate a loop. After the pure screw triple junction is formed, out-of-plane pure shear is applied to one of its possible cross slip planes, the $(1\bar{1}0)$ plane,
%
%
and the motion of the $1/2[111]$ screw dislocation is limited to this plane. 

\begin{figure}[!ht]
\centering\includegraphics[width=0.9\linewidth]{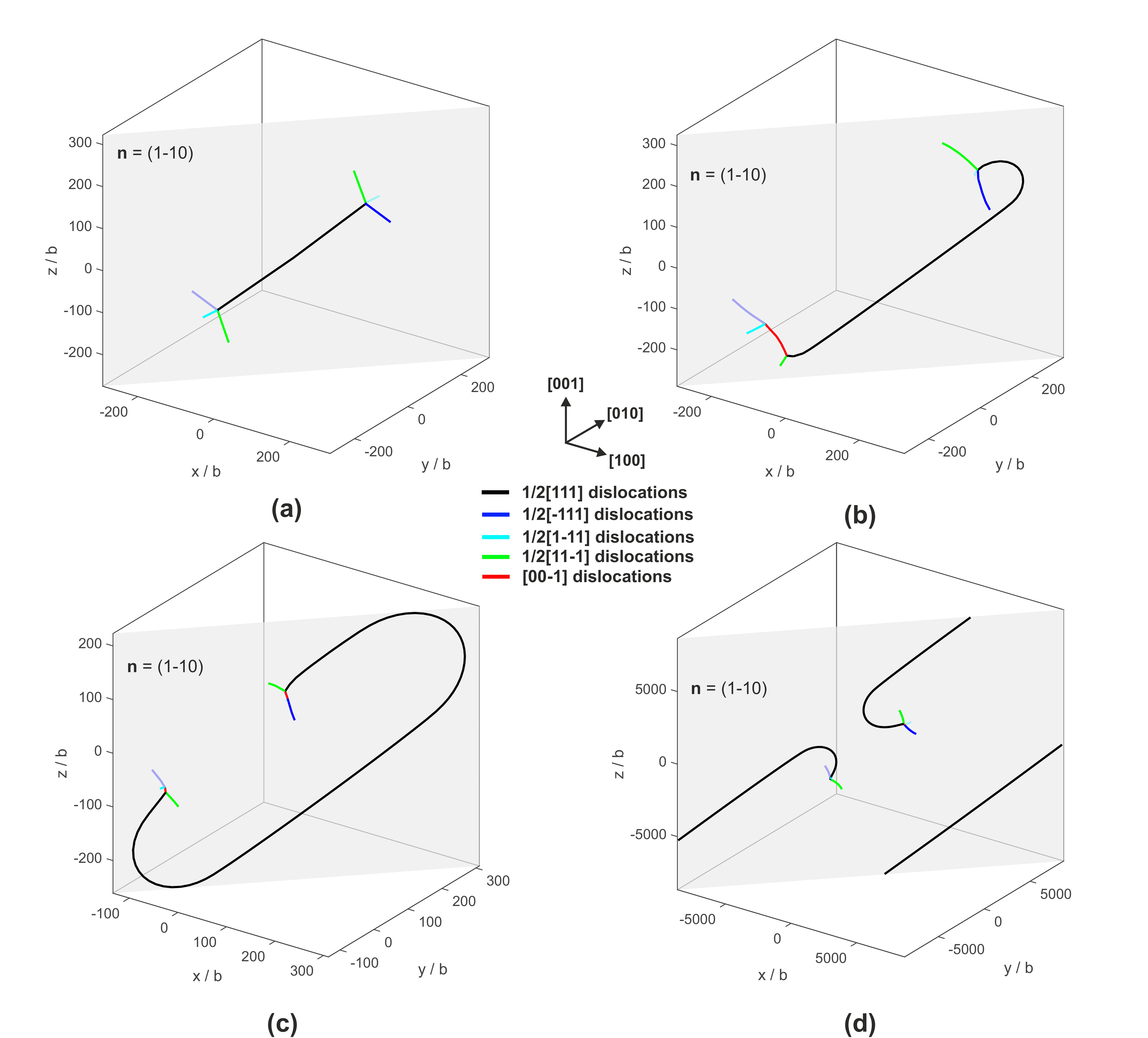}
\caption{A screw triple junction operating as a F-R source. (a) shows the initial triple junction before bow-out; (b) shows the interaction between the $1/2[111]$ (black) dislocation loop and the $1/2[11\bar{1}]$ (green) arm on the left, forming a $[00\bar{1}]$ sessile dislocation (red): $1/2[11\bar{1}](1\bar{1}0)+1/2[\bar{1}\bar{1}\bar{1}](1\bar{1}0)\rightarrow [00\bar{1}](1\bar{1}0)$; (c) the $1/2[111]$ loop has bowed back restoring the $1/2[11\bar{1}]$ arm and the process then occurs on the right. (d) Shows the moment just before the emission of a full loop, restoring the triple junction as in (a) and the process repeats.}
\label{fig:tripleFRprocess}
\end{figure}

When the applied stress is sufficiently high, the triple junction will bow out and generate dislocation loops, as shown in \autoref{fig:tripleFRprocess}. 
This procedure is similar to the F-R operation but has some unique features which haven't been discussed before. On each of the possible cross slip planes (i.e. the possible F-R operation planes), there lie a pair of $1/2\langle111\rangle$ arms formed during the relaxation process. The bowing out loop will inevitably interact with one of these pairs during the operation. In the case selected here, it is the $1/2[11\bar{1}]$ (green) pair. As shown in \autoref{fig:tripleFRprocess}(b), the junction encounters the arm on the left as soon as it bows out, which yields a sessile $[00\bar{1}]$ segment (red): $1/2[11\bar{1}](1\bar{1}0)+1/2[\bar{1}\bar{1}\bar{1}](1\bar{1}0)\rightarrow [00\bar{1}](1\bar{1}0)$. Note that the line direction of the junction is opposite to the $1/2[11\bar{1}]$ arm when they first interact. This impedes the expansion of the loop on the left and leads to the unsymmetric development of the loop as shown in \autoref{fig:tripleFRprocess}(b). As expansion continues, the loop will eventually bypass the pinning point, bow sharply back and wrap around the now sessile arm (red) restoring it back to it's initial state (green): $[00\bar{1}](1\bar{1}0)+1/2[111](1\bar{1}0)\rightarrow 1/2[11\bar{1}](1\bar{1}0)$. The same process, where a tempory sessile  $[00\bar{1}]$ segment is formed, then occurs on the other pinning point; as shown on the right of \autoref{fig:tripleFRprocess}(c). 

\begin{figure}[!ht]
\centering\includegraphics[width=0.9\linewidth]{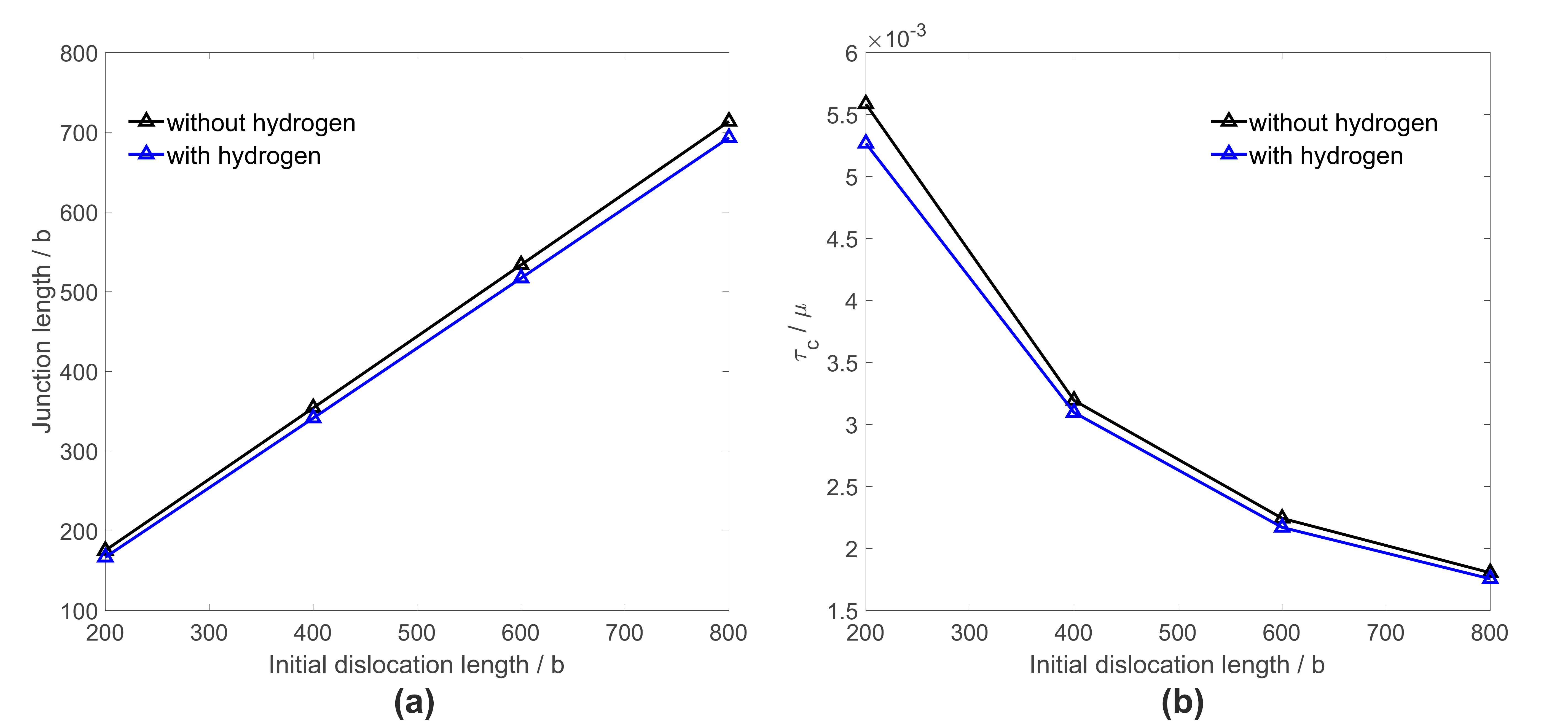}
\caption{Variation of (a) triple junction length and (b) strength with initial dislocation length, with and without hydrogen.}
\label{fig:triplestrength}
\end{figure}

During the activation of a F-R source and similarly of the triple junction, there exists a certain geometric configuration beyond which the dislocation line becomes unstable and keeps growing and emits loops. During the simulation, the magnitude of the applied stress is gradually increased until the triple junction starts to operate and reaches the critical geometry \citep{Cai2018}, and thus the critical activation stress $\tau_c$, i.e. junction strength, is determined. We have verified that the predicted activation stresses obtained for pure screw F-R sources in the absence of hydrogen agree with the $\frac{3}{2}\mu b/L$ relation expected by theoretical calculation \citep{anderson2017}.
The formation length and destruction strength of the triple junctions of varying length with and without hydrogen ($c_0=10.0$appm) are shown in \autoref{fig:triplestrength}.
Consistent with the results in \autoref{fig:DDDlengths}(b), the hydrogen core force slightly reduces the length and strength of the triple junction. However the effect is weak as it is a second order effect. 

\begin{figure}[!ht]
\centering\includegraphics[width=0.9\linewidth]{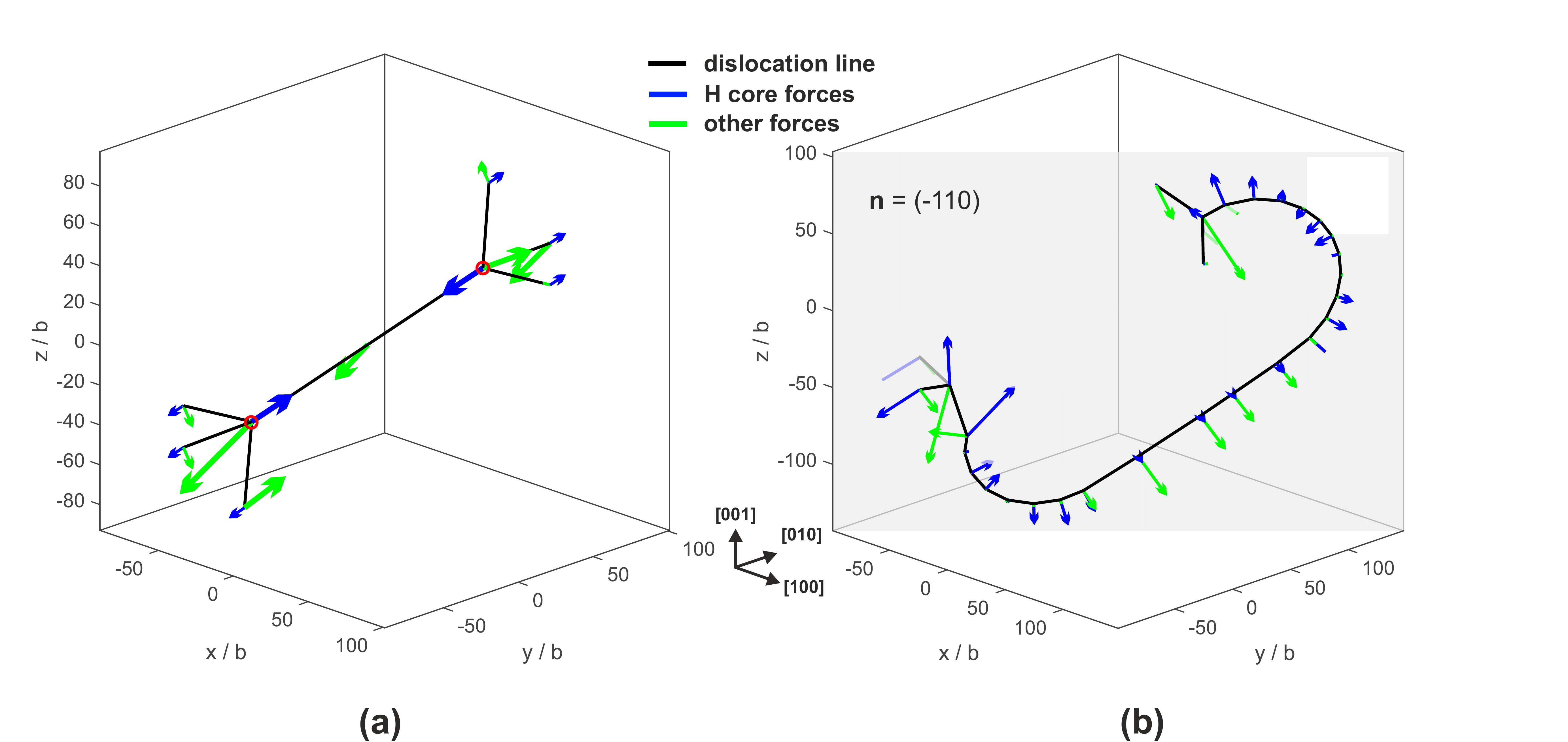}
\caption{The nodal force partitioned into hydrogen core force, ${\bm{F}}^{c,H}$ (blue arrows), and non-hydrogen parts $\tilde{\bm{F}}_k + \hat{\bm{F}}_k$ (green arrows). (a) shows the formation of the triple junction. (b) shows the bow-out of the junction. The ending points of the junction in (a) are marked with red circles; note ${\bm{F}}^{c,H}$ has been scaled $20\times$ with respect to $\tilde{\bm{F}}_k + \hat{\bm{F}}_k$ for clarity.}
\label{fig:tripleforce}
\end{figure}

The nodal forces are shown in \autoref{fig:tripleforce}.
In \autoref{fig:tripleforce}(a), the hydrogen core force tends to shrink the triple junction. During the activation of the triple junction, as shown in \autoref{fig:tripleforce}(b), the hydrogen core force tends to enhance bowing out of the mixed dislocation segments while impeding the edge segments, which facilitates the bow-out and nucleation of a loop. This accounts for the slight reduction in stress with hydrogen in \autoref{fig:triplestrength}(b). A hydrogen dependent mobility law \citep{Yu2019a} was not incorporated in the current simulations. A significant enhancement in the operation of the triple junction/F-R source is expected if hydrogen enhanced screw mobility is implemented.

It is hard to anticipate the consequence of hydrogen on the mechanical response by analysis of individual junctions. For this purpose, hydrogen-junction interactions should be discussed in the context of a complex dislocation network formed in a volume of material under an applied load.
\subsection{Microcantilever simulation with hydrogen core force shielding}
\label{subsec:microcantilever}

To examine if hydrogen core force shielding could effect mechanical properties we performed DDD simulations of a microcantilever bend test. The beam is aligned along the $\langle 100 \rangle$ crystallographic axes and has dimensions of $12\times3\times3~\mu\textrm{m}$ with $\bm{u}(0,y,z)=[0,0,0]$,  $\bm{u}(L_x,y,L_z) = [0,0,U]$ with an applied displacement of $U=-0.08~\mu\textrm{m}$, and the remaining surfaces are traction free. Fully integrated linear brick finite elements, $0.24\mu \textrm{m}$ in size, were used to solve for the corrective elastic fields at every time increment \citep{Yu2019a}. The initial dislocation structure consisted of $40$ square prismatic loops with a length of $0.39\mu \textrm{m}$ randomly generated from the three slip systems of interest in \autoref{eq:triple}, $1/2[\bar{1}11](01\bar{1})$, $1/2[1\bar{1}1](10\bar{1})$ and $1/2[11\bar{1}](1\bar{1}0)$, positioned randomly in the high stress region near the fixed end $x<4~\mu\textrm{m}$. This enables the formation of binary junctions as well as triple and higher order junctions referred to here as multi junctions. A hydrogen concentration of $10$ appm was applied when computing the core force, apart from this the simulations with and without hydrogen were identical.
\begin{figure}[!ht]
\centering\includegraphics[width=0.95\linewidth]{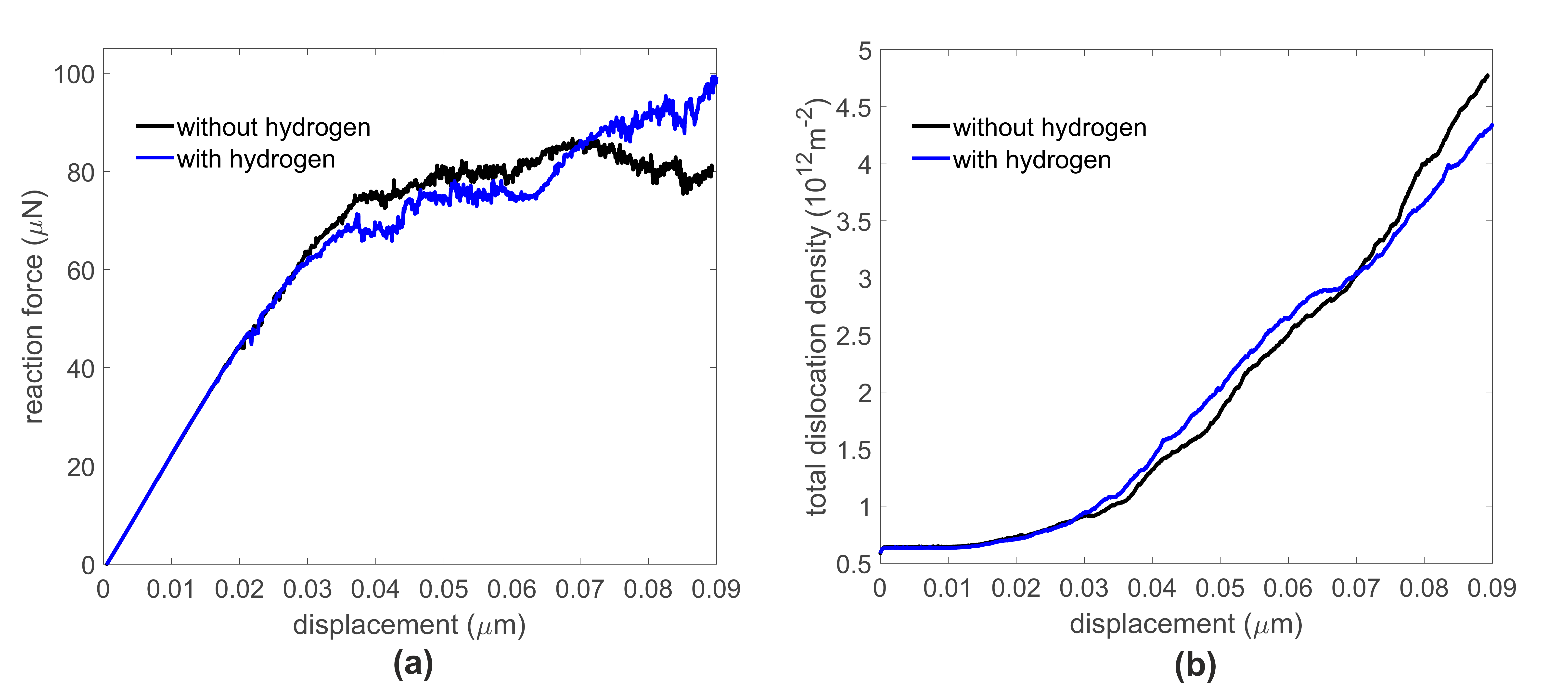}
\caption{(a) The load-displacement curve; (b) the evolution of the total dislocation density.}
\label{fig:load-disp}
\end{figure}
The evolution of the total dislocation density, together with the load-displacement curves are plotted in \autoref{fig:load-disp}. During the early loading stage ($0.03<U<0.07~\mu\textrm{m}$), the dislocation density is slightly higher in the presence of hydrogen, due to hydrogen promoted dislocation generation from the prismatic loops, due to a similar mechanism as illustrated in \autoref{fig:tripleforce}(b). Consequently, yielding occurs earlier with hydrogen.
\begin{figure}[!ht]
\centering\includegraphics[width=0.95\linewidth]{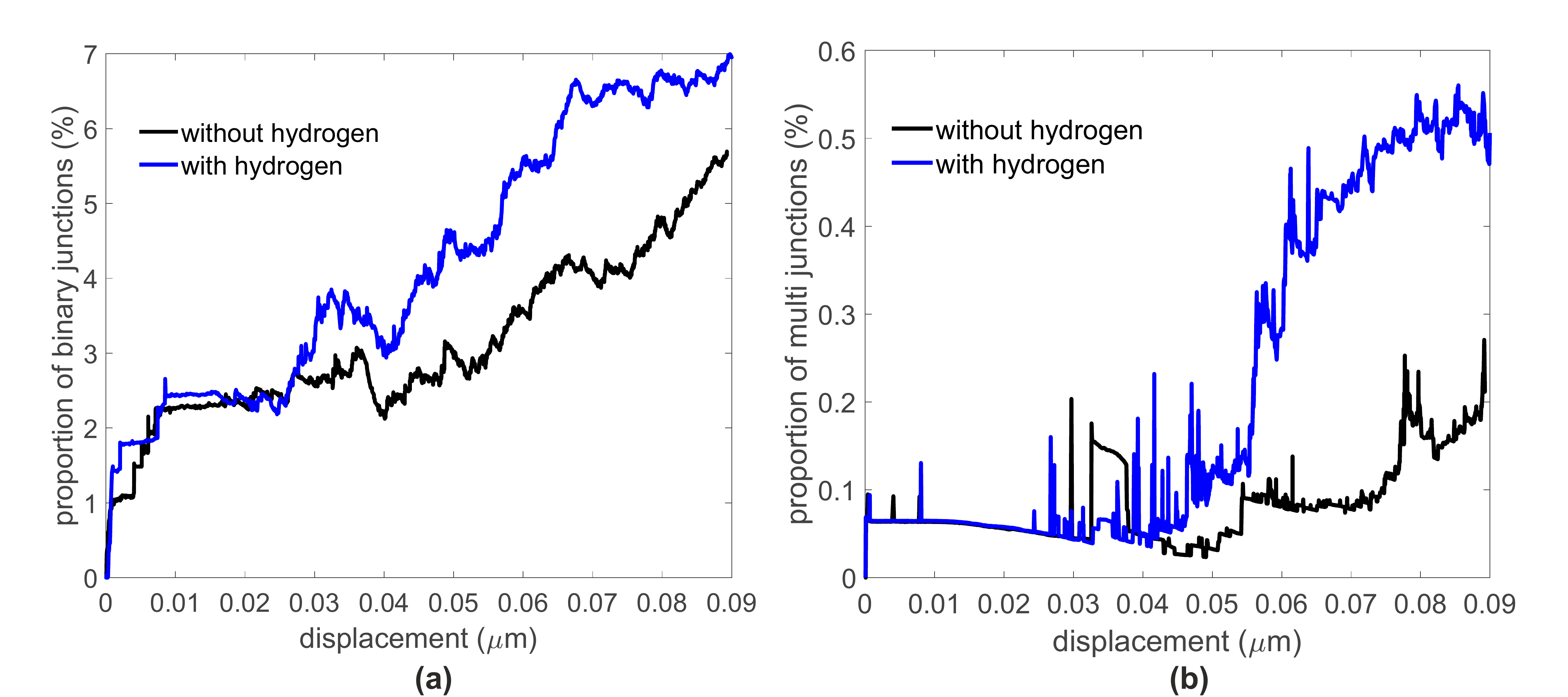}
\caption{(a) The evolution of binary dislocation junction density; (b) the evolution of multiple dislocation junction density.}
\label{fig:junctiondensity}
\end{figure}

The evolution of the binary and multi junction densities with and without hydrogen are shown in \autoref{fig:junctiondensity}. During the entire loading history, binary junctions are the dominant junction type. The density of binary junctions is significantly higher in the presence of hydrogen, as shown in \autoref{fig:junctiondensity}(a). This is consistent with hydrogen increasing the length and strength of binary junctions as discussed earlier. In \autoref{fig:junctiondensity}(b), hydrogen also increases the proportion of multi junctions, which usually form when a mobile dislocation interacts with a binary junction to form a triple junction.

\begin{figure}[!ht]
\centering\includegraphics[width=0.9\linewidth]{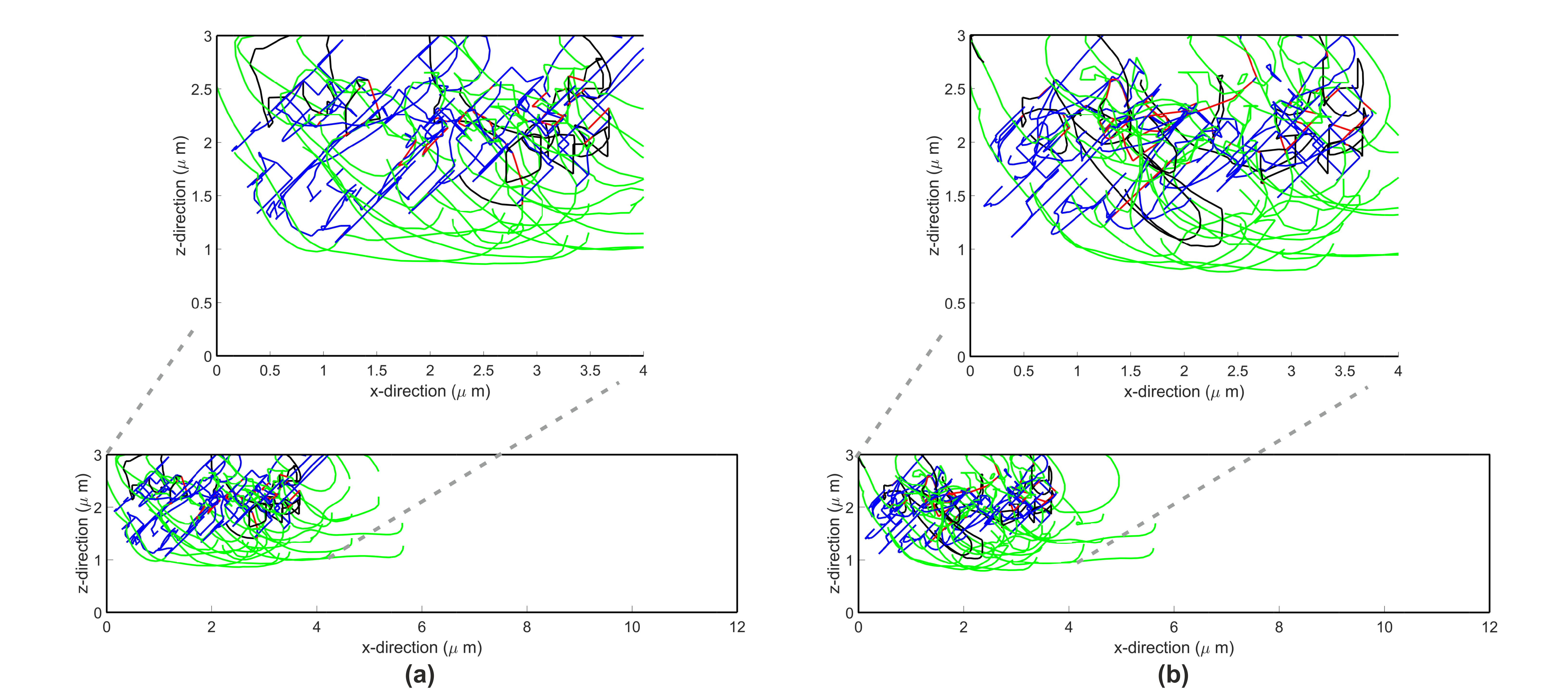}
\caption{Dislocation structure in the initial loading stage $U=-0.0524~\mu\textrm{m}$: (a) without hydrogen and (b) with hydrogen. The dislocations belonging to the three initial slip systems are $1/2[\bar{1}11](01\bar{1})$ (black), $1/2[1\bar{1}1](10\bar{1})$ (blue) and $1/2[11\bar{1}](1\bar{1}0)$ (green) with binary junctions (red) and multiple dislocation junctions (magenta).}
\label{fig:dislocationstruct}
\end{figure}

The increased junction density with hydrogen eventually reduces the mobile density production rate leading to subsequent hardening, as evident in \autoref{fig:load-disp}(a) for $U>0.07 ~\mu\textrm{m}$. This corresponds to the mobile density with hydrogen dropping below that without hydrogen. A snap shot (at $U=-0.0524\mu m$) of the dislocation structure in the initial (hydrogen softening) stage is shown in \autoref{fig:dislocationstruct}.
The $1/2[1\bar{1}1](10\bar{1})$ and $1/2[11\bar{1}](1\bar{1}0)$ slip systems have the highest Schmid factor and so are the most active and make up the majority of the mobile density. The higher binary junction density (red) with hydrogen is also visible in \autoref{fig:dislocationstruct}(b).

\begin{figure}[!ht]
\centering\includegraphics[width=0.9\linewidth]{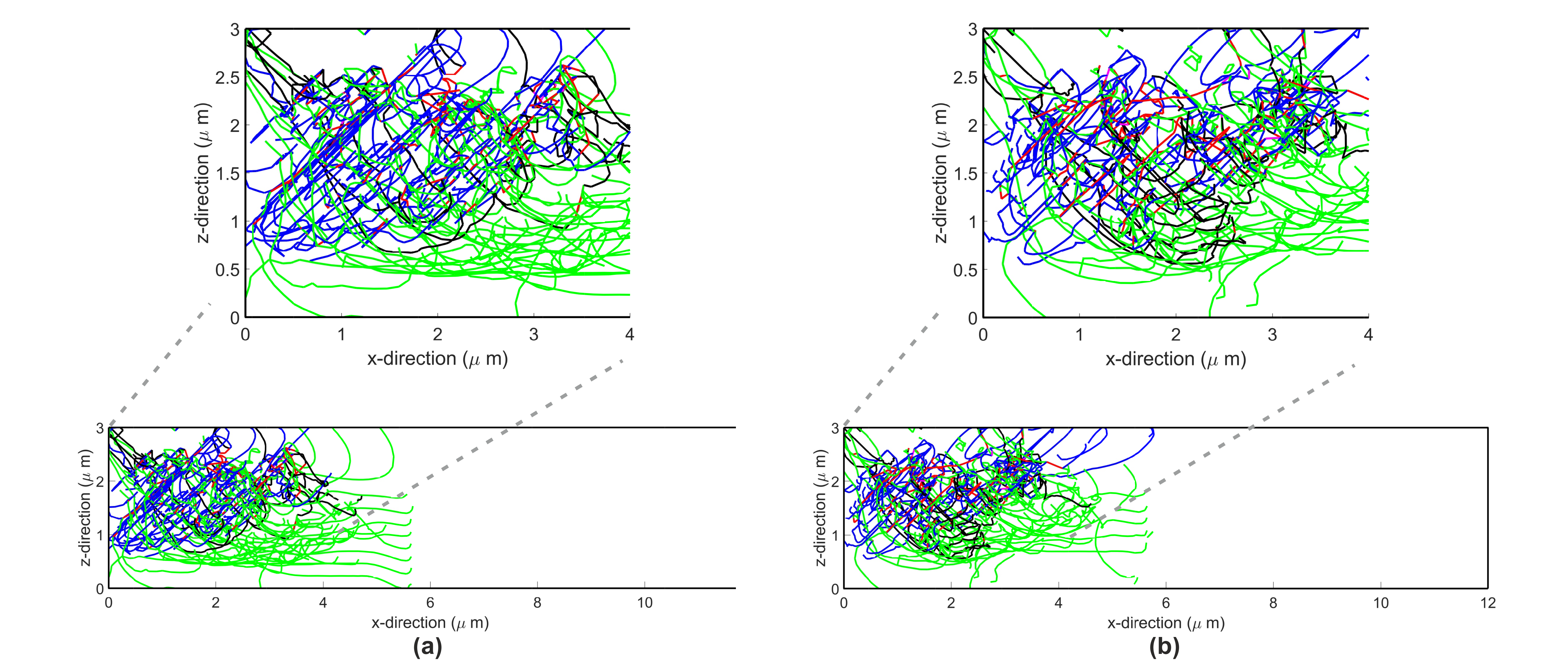}
\caption{Dislocation structure at $U=-0.09\mu m$: (a) without hydrogen and (b) with hydrogen. With the same colour coding as \autoref{fig:dislocationstruct}.}
\label{fig:dislocationstruct2}
\end{figure}

The dislocation structure at the end of the simulation ($U=-0.09\mu m$) is shown in \autoref{fig:dislocationstruct2}
Comparing \autoref{fig:dislocationstruct2}(a) with \autoref{fig:dislocationstruct}(a) shows an increase in the mobile density (green and blue) in the absence of hydrogen. Comparing \autoref{fig:dislocationstruct2}(b) with \autoref{fig:dislocationstruct2}(a) shows the higher junction density and lower mobile density with hydrogen.

Hydrogen significantly increases the density of junctions, leading to more complicated dislocation tangles, which will inevitably impede the motion and generation of dislocations as loading increases. Therefore, the overall influence of hydrogen core force shielding is expected to be a slight softening effect initially followed by a hardening effect. Recently, \cite{Wang2019} observed that hydrogen decreases the yield stress and increases strain hardening in an fcc TWIP steel and hydrogen hardening was observed in bcc iron, both experimentally \citep{Depover2016} and numerically \citep{Xie2011}. 
\section{Summary}
\label{sec:sum}
The hydrogen core force was calibrated using atomistic modelling and shows that hydrogen reduces or shields the core force of edge dislocations. This effect plays a dominant role in the formation of dislocation junctions even at low concentrations, where the contribution from hydrogen elastic shielding is negligible.

Hydrogen was found to increase the length and strength of $\langle100\rangle$ binary edge junctions whereas the influence on a $1/2\langle111\rangle$ screw triple junctions was negligible. \textcolor{black}{However the increased binary junction strength leads to a significant increase in the number of triple junctions forming in microscale plasticity simulations}. The triple junction can operate like a F-R source, hydrogen facilitates this by exerting a core force promoting the bowing-out of the junction and enhancing the expansion of the nucleated loops. By strengthening binary junctions, hydrogen tends to limit the motion and generation of dislocations; while it tends to accelerate dislocation generation from the triple junctions.

Hydrogen was found to increase the total dislocation density and cause plastic softening at an early stage in a microcantilever. Hydrogen was also observed to increase the density of binary and multi junctions. Junctions act as obstacles to dislocation motion. Consequently, dislocation generation and motion were suppressed by hydrogen as the load increased, enhancing strain hardening. \textcolor{black}{The hydrogen core force shielding effect analysed here appears to share many similarities with the hydrogen elastic shielding effect but occurs at much lower concentrations as typically found in bcc materials.}

\section*{Acknowledgements}
This work was supported by the Engineering and Physical Sciences Research Council (EPSRC) under Programme grant EP/L014742/1 and Fellowship grant EP/N007239/1; IHK acknowledges support from Bulgaria National Science Fund (BNSF) under Programme grant KP-06-H27/19; HY acknowledges Bruce Bromage for useful discussions.




\section*{References}
\bibliographystyle{model2-names}
\bibliography{Hcore.bib}







\end{document}